\documentclass[prd,showpacs,twocolumn,preprintnumbers,amsmath,amssymb,superscriptaddress,floatfix,nofootinbib]{revtex4-2}%

\usepackage{graphicx}
\usepackage{bm}
\usepackage{amsmath}
\usepackage{amsfonts}
\usepackage{amssymb}
\usepackage{color}
\usepackage{multirow}
\usepackage{soul}
\usepackage{subfigure}
\usepackage[colorlinks, citecolor=blue,anchorcolor=red,menucolor=red, linkcolor=red,filecolor=red,runcolor=red,urlcolor=blue,frenchlinks=red]{hyperref}

\begin{document}
	\title{The $B^-\to J/\psi\eta^\prime K^-$ reaction and the $Y(4710)$ and $K_0^*(1430)$ contributions}
	
\author{Wen-Tao Lyu}
\affiliation{School of Physics, Zhengzhou University, Zhengzhou 450001, China}

\author{Man-Yu Duan}\email{duanmy@seu.edu.cn}
\affiliation{School of Physics, Southeast University, Nanjing 210094, China}
\affiliation{School of Physics, Zhengzhou University, Zhengzhou 450001, China}
\affiliation{Departamento de Física Teórica and IFIC, Centro Mixto Universidad de Valencia-CSIC Institutos de Investigación de Paterna, 46071 Valencia, Spain}

\author{De-Min Li}\email{lidm@zzu.edu.cn}
\affiliation{School of Physics, Zhengzhou University, Zhengzhou 450001, China}

\author{Bing Wang}\email{bingwang@zzu.edu.cn}
\affiliation{School of Physics, Zhengzhou University, Zhengzhou 450001, China}

\author{Dian-Yong Chen}\email{chendy@seu.edu.cn}
\affiliation{School of Physics, Southeast University, Nanjing 210094, China}
\affiliation{Lanzhou Center for Theoretical Physics, Lanzhou University, Lanzhou 730000, P. R. China}

\author{En Wang}\email{wangen@zzu.edu.cn}
\affiliation{School of Physics, Zhengzhou University, Zhengzhou 450001, China}

\begin{abstract}
Motivated by the recent LHCb Collaboration analysis, we have investigated the process $B^-\to J/\psi\eta^\prime K^-$ by considering the contributions from the resonances $Y(4710)$ and $K_0^*(1430)$. Our results are in good agreement with the measured invariant mass distributions and the Dalitz plot of LHCb, which supports the existence of the $K_0^*(1430)$ and the $Y(4710)$ in the process $B^-\to J/\psi\eta^\prime K^-$. 
Since the decay mode of $Y(4710)\to J/\psi \eta'$ is Okubo-Zweig-Iizuka (OZI) suppressed for the charmonium state, the measurements of the $\mathcal{B}(Y(4710)\to J/\psi\eta')$ through the process $B^-\to J/\psi \eta' K^-$ or other processes is helpful to understanding the nature of the $Y(4710)$. Thus, we advocate that Belle~II and LHCb Collaborations could perform the more precise analysis to confirm the evidence of the $Y(4710)$ state, which could be helpful to reduce the experimental uncertainties of its mass and width, and to explore its nature.


\end{abstract}
	
	\pacs{}
	\date{\today}
	
	\maketitle
	
\section{Introduction}\label{sec1}
Hadron states, the subatomic particles bound by and interacting through the strong force, are classified as mesons of quark-antiquark pair and baryons of three quarks within the conventional quark model. Quantum Chromodynamics (QCD), the quantum field theory of the strong force, has provided many high precision predictions tested for high energies, but fail to directly access the properties of the hadrons, due to the fact that the QCD coupling constant is large for low energies~\cite{Mai:2025wjb}. The exotic hadrons, such as hybrids, pentaquark states, tetraquark states, glueballs, are expected to exist within QCD~\cite{Klempt:2007cp}, and many candidates of the exotic hadrons were observed in experiments in last two decades, which has deepened our understanding of the properties of the QCD in low energies~\cite{Lyu:2025oow,Guo:2017jvc,Guo:2019twa,Lu:2024dtb,Wang:2024jyk}.

Since the $X(3872)$ has been observed by the Belle Collaboration in 2003~\cite{Belle:2003nnu}, many charmonium-like states have been reported experimentally, and called many attentions~\cite{Chen:2016qju,Liu:2019zoy,Brambilla:2004jw,Chen:2022asf}. In 2023, the BESIII Collaboration observed a new enhancement structure $Y(4710)$ with a mass of $M=(4704.0\pm52.3\pm69.5)$~MeV and a width of $\Gamma=(183.2\pm114.0\pm96.1)$~MeV in the cross sections for the process $e^+e^-\to K_S^0K_S^0J/\psi$, and the statistical significance is $4.2\sigma$~\cite{BESIII:2022kcv}.
Subsequently, using the data samples with an integrated luminosity of 5.85~$fb^{-1}$ collected at center-of-mass energies from 4.61 to 4.95~GeV, BESIII has measured the cross sections for the process $e^+e^-\to K^+K^-J/\psi$, and confirmed the resonance $Y(4710)$ with a significance over $5\sigma$~\cite{BESIII:2023wqy}. Its mass and width are measured to be $M=(4708^{+17}_{-15}\pm21)$~MeV, $\Gamma=(126^{+27}_{-23}\pm30)$~MeV, respectively~\cite{BESIII:2023wqy}, which are  consistent with the previous measurements of BESIII within the experimental uncertainties~\cite{BESIII:2022kcv}.

Although the mass of $Y(4710)$ observed by BESIII is close to the $\psi(5S)$ mass (4704~MeV or 4711~MeV) predicted by the linear potential quark model~\cite{Ding:2007rg,Deng:2016stx,Gui:2018rvv}, there are also different theoretical explanations for the structure of the $Y(4710)$ state~\cite{Brambilla:2022hhi,Nakamura:2023obk,Chen:2023oqs}. In Refs.~\cite{Wang:2023jaw,Wang:2025sic}, the authors deem that the $Y(4710)$ is a good candidate of the hidden-charm-hidden-strange tetraquark state $[sc]_{\tilde{V}}[\bar{sc}]_A-[sc]_A[\bar{sc}]_{\tilde{V}}$ with $J^{PC}=1^{--}$ within the framework of the QCD sum rules. In Ref.~\cite{Deng:2023mza}, the authors have studied the charmonium spectrum within an unquenched quark model including coupled channel effects, and found that the $Y(4710)$ can be interpreted as $\psi(4D)$ state. In addition, the $Y(4710)$  can also be interpreted as the hybrid candidates of $H_1(4507)$~\cite{Berwein:2024ztx} or $H_1[1^{--}](4812)$~\cite{Brambilla:2022hhi}. It is notable that the mass and width of the $Y(4710)$ have very large uncertainties, and this state has not yet been listed in Review of Particle Physics~(RPP)~\cite{ParticleDataGroup:2024cfk}, which implies that the $Y(4710)$ is not well-understood, and still needs confirmation in more processes.

The hadronic decays of the bottom hadrons provide an ideal platform to investigate the charmonium-like states~\cite{Brambilla:2019esw,Li:2023nsw,Duan:2023qsg,Wang:2024gsh,Wu:2023rrp,Wang:2022xga,Wei:2021usz,Liu:2020ajv,Zhang:2020rqr,Wang:2017mrt,Dai:2018nmw}. 
Recently, the LHCb Collaboration has observed the process $B^+\to J/\psi\eta^\prime K^+$~\cite{LHCb:2023qca}, and given $\mathcal{B}(B^+\to J/\psi \eta' K^+)=(3.06\pm 0.29\pm 0.18\pm 0.04)\times 10^{-5}$. 
From Fig.~2(b) of Ref.~\cite{LHCb:2023qca}, one can find a peak structure around 4.7~GeV in the $J/\psi \eta'$ invariant mass distribution, close to the mass of $Y(4710)$. If this peak structure can  be associated with $Y(4710)$, the decay mode of $Y(4710)\to J/\psi \eta'$, which is Okubo-Zweig-Iizuka (OZI) suppressed for the charmonium state decay~\cite{Okubo:1963fa,Zweig:1964ruk,Iizuka:1966fk}\footnote{In Ref.~\cite{Iizuka:1966fk}, the selection principle is `{\it Among all the possible effective vertices of a given process, the dominant one corresponds to the connected diagram viewed from the composite mode}.' The process $Y(4710)(c\bar{c})\to J/\psi \eta'$ is analogous to the disconnected process of Fig.~2(iii) of Ref.~\cite{Iizuka:1966fk}.}, could be helpful to explore its nature, and the precise measurements of this process could be useful to determine the accurate properties of the $Y(4710)$.

Moreover, a near-threshold enhancement structure also appears in the $\eta' K^+$ invariant mass distribution, which could associated with the excited kaon states. According to RPP~\cite{ParticleDataGroup:2024cfk}, there are several excite kaon states around 1.4~GeV, the $K_1(1400)$ ($J^P=1^+$), $K^*(1410)$ ($J^P=1^-$), $K^*_0(1430)$ ($J^P=0^+$), $K^*_2(1430)$ ($J^P=2^+$), and $K(1460)$ ($J^P=0^-$). The states $K_1(1400)$ and $K(1460)$ can not couple to $\eta'K $ because of the conservation of the parity and angular momentum, and the decay modes of $K^*(1410)\to \eta'K$ and $K^*_2(1430)\to \eta'K$ are not observed in experiments~\cite{ParticleDataGroup:2024cfk}. Indeed, only the $K^*_0(1430)$ is observed to couple to $\eta'K $ mode, such as in $\chi_{cJ} \to \eta^{\prime}K^+K^-$ by BESIII~\cite{BESIII:2014dlb} and $\eta_c\to \eta^\prime K^+K^-$ by \textit{BABAR}~\cite{BaBar:2021fkz}. 
On the theoretical side, the $K_0^*(1430)$ has call many theoretical studies~\cite{Ikeno:2024fjr,Lyu:2024wxa,Li:2023nsw}. Meanwhile, the $K_0^*(1430)$ could be explained as the dynamically generated state from the vector meson-vector meson interaction~\cite{Abreu:2023yvf,Geng:2008gx,Dai:2023jix,Garcia-Recio:2013uva}, which will be adopted to present the contribution of the intermediate $K_0^*(1430)$.




In the present work, we will investigate the process $B^-\to J/\psi\eta^\prime K^-$\footnote{In this work, we study the process $B^-\to J/\psi\eta^\prime K^-$, the charge conjugation process of the $B^+\to J/\psi\eta^\prime K^+$, because it is convenient to deal with the weak decay of the $b$ quark.} by considering the contributions from the resonances $K_0^*(1430)$ and  $Y(4710)$, respectively, and try to discuss whether the $Y(4710)$ can be produced in the process $B^-\to J/\psi\eta^\prime K^-$ or not according to the LHCb measurements~\cite{LHCb:2023qca}, which should be an incentive for the more accurate experimental analysis.

This paper is organized as follows. In Sec.~\ref{sec2}, we present the theoretical formalism of the process $B^-\to J/\psi\eta^\prime K^-$. Numerical results and discussion are shown in Sec.~\ref{sec3}. Finally, we give a short summary in the last section.

\section{Formalism}\label{sec2}

In this section, we will introduce the theoretical formalism adopted in the present estimations. The mechanisms for the intermediate state $K_0^*(1430)$ is given in Sec.~\ref{sec2a}, and the mechanism of the intermediate state $Y(4710)$ is given in Sec.~\ref{sec2b}. Finally, we give the formalism for the invariant mass distributions for the process $B^-\to J/\psi\eta^\prime K^-$ in Sec.~\ref{sec2c}.

\subsection{Contribution of $K_0^*(1430)$ in the $B^-\to J/\psi\eta^\prime K^-$}\label{sec2a}

\begin{figure}[htbp]
	\centering
	
	\includegraphics[scale=0.65]{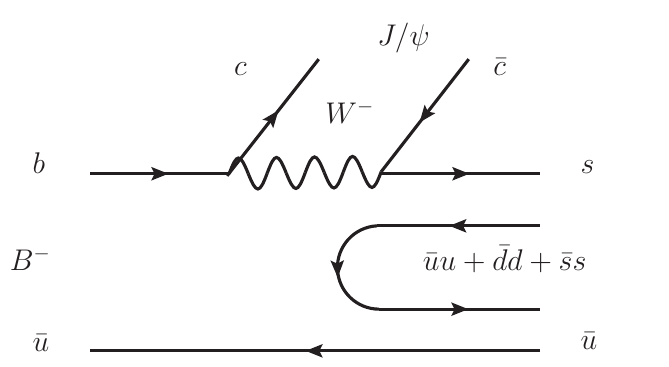}
	
	\caption{ Quark level diagram for the process $B^-\to J/\psi  s(\bar{u}u+\bar{d}d+\bar{s}s)\bar{u}$.}\label{fig:MM-quark}
\end{figure}

Here, we will consider the contribution from the $S$-wave vector meson-vector meson final state interaction and the $B^-\to J/\psi VV\to J/\psi\eta^\prime K^-$ reaction. As shown in Fig.~\ref{fig:MM-quark}, the $b$ quark of the initial $B^-$ weakly decays into a $c$ quark and a $W^-$ boson, then the $W^-$ boson subsequently decays into a $\bar{c}$ quark and an $s$ quark. The $c$ quark and $\bar{c}$ from the $W^-$ boson will hadronize into the $J/\psi$ meson. The $s$ quark and the $\bar{u}$ quark from the initial $B^-$, together with the quark pair $\bar{q}q=\bar{u}u+\bar{d}d+\bar{s}s$ created from the vacuum with the quantum numbers $J^{PC}=0^{++}$, hadronize into hadron pairs, as follows,\footnote{Here, the produced light quark pairs also can hadronize into other meson pairs, however it is shown that the $K_0^*(1430)$ could be explained as the dynamically generated state from the vector meson-vector meson interaction in Refs.~\cite{Abreu:2023yvf,Dai:2023jix,Geng:2008gx,Garcia-Recio:2013uva}. Therefore, we have only considered the hadronization of the vector-vector meson pairs here.}
\begin{equation}
	\sum_is(\bar{u}u+\bar{d}d+\bar{s}s)\bar{u} = \sum_{i=1}^3V_{3i}V_{i1},
\end{equation}
where $i=1,2,3$ correspond to the $u$, $d$, and $s$ quarks, respectively, and $V$ is the $3\times 3$ matrix of the vector mesons~\cite{Zhang:2024myn},
\begin{equation}
	V=\left(\begin{array}{ccc}
		\frac{\omega+\rho^0}{\sqrt{2}} & \rho^{+} & K^{*+} \\
		\rho^{-} & \frac{\omega-\rho^0}{\sqrt{2}} & K^{* 0} \\
		K^{*-} & \bar{K}^{* 0} & \phi
	\end{array}\right).
\end{equation}

Analytically, we have
\begin{equation}\label{eq:h_factor}
		\begin{aligned}
			B^-&=b\bar{u}\nonumber\\
			&\Rightarrow cW^-\bar{u}\nonumber\\
			&\Rightarrow c\bar{c}s\bar{u}\nonumber\\
			&=J/\psi \sum_is\bar{q}_iq_i\bar{u} \nonumber\\
			&=J/\psi \sum_iV_{3i}V_{i1} \nonumber\\
			&=J/\psi \left\{K^{*-}\left(\frac{\rho^0}{\sqrt{2}}+\frac{\omega}{\sqrt{2}}\right)+\bar{K}^{*0}\rho^-+K^{*-}\phi\right\} \nonumber\\
			&=J/\psi \left\{\frac{1}{\sqrt{2}}K^{*-}\rho^0+\frac{1}{\sqrt{2}}K^{*-}\omega+\bar{K}^{*0}\rho^-+K^{*-}\phi\right\}.
		\end{aligned}
		\end{equation}

With the isospin multiplets of $(K^{*+},K^{*0})$, $(\bar{K}^{*0},-K^{*-})$, and $(-\rho^+, \rho^0, \rho^-)$~\cite{Close:1979bt,Li:2024rqb,Li:2024tvo,Duan:2020vye}, we have,
\begin{eqnarray}
	\left|K^{*-}\rho^0\right>&=&-\left|\frac{1}{2},-\frac{1}{2}\right>\left|1,0\right> \nonumber\\
	&=&-\left[\sqrt{\frac{2}{3}}\left|\bar{K}^{*}\rho,\frac{3}{2},-\frac{1}{2}\right>-\sqrt{\frac{1}{3}}\left|\bar{K}^{*}\rho,\frac{1}{2},-\frac{1}{2}\right>\right]\nonumber\\
	&=&-\sqrt{\frac{2}{3}}\left|\bar{K}^{*}\rho,\frac{3}{2},-\frac{1}{2}\right>+\sqrt{\frac{1}{3}}\left|\bar{K}^{*}\rho,\frac{1}{2},-\frac{1}{2}\right>,
\end{eqnarray}
\begin{eqnarray}
	\left|\bar{K}^{*0}\rho^-\right>&=&\left|\frac{1}{2},\frac{1}{2}\right>\left|1,-1\right> \nonumber\\
	&=&\sqrt{\frac{1}{3}}\left|\bar{K}^{*}\rho,\frac{3}{2},-\frac{1}{2}\right>+\sqrt{\frac{2}{3}}\left|\bar{K}^{*}\rho,\frac{1}{2},-\frac{1}{2}\right>,
\end{eqnarray}
\begin{eqnarray}
	\frac{1}{\sqrt{2}}\left|K^{*-}\rho^0\right>+\left|\bar{K}^{*0}\rho^-\right>&=&\left(\sqrt{\frac{1}{6}}+\sqrt{\frac{2}{3}}\right)\left|\bar{K}^{*}\rho,\frac{1}{2},-\frac{1}{2}\right> \nonumber\\
	&=&\frac{3}{\sqrt{6}}\left|\bar{K}^{*}\rho,\frac{1}{2},-\frac{1}{2}\right>.
\end{eqnarray}
In the isospin basis, we can obtain the $\bar{K}^{*}\rho(I=1/2)$, $\bar{K}^{*}\omega(I=1/2)$, and $\bar{K}^{*}\phi(I=1/2)$ channels.
\begin{equation}
	H=J/\psi \left\{\frac{3}{\sqrt{6}}\bar{K}^{*}\rho+\frac{1}{\sqrt{2}}\bar{K}^{*}\omega+\bar{K}^{*}\phi\right\}. \label{eq:hadronize}
\end{equation}
\begin{figure}[htbp]
	\centering
	
	\includegraphics[scale=0.65]{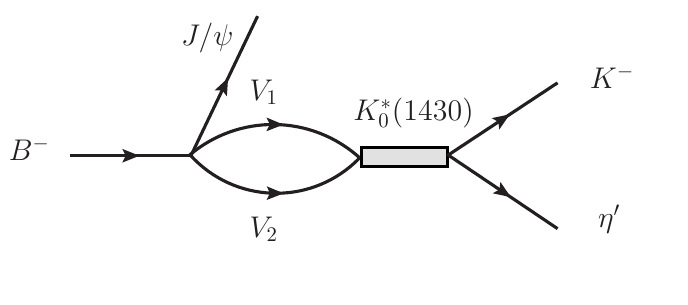}
	
	\caption{Mechanism for $K^*_0(1430)$ state.}\label{fig:K1430-hardon}
\end{figure}

This said, diagrammatically we have the mechanisms depicted in Fig.~\ref{fig:K1430-hardon}, which could produce the $K^*_0(1430)$, and the amplitude could be expressed as,
\begin{equation}
	\mathcal{T}^{K^*_0(1430)}=p_{J/\psi}\sum_ih_i\tilde{G}_i\dfrac{g_ig_{\eta^{\prime}K^-}}{M_{\eta^{\prime}K^-}^2-M_{K^*_0}^2+iM_{K^*_0}\Gamma_{K^*_0}}, \label{eq:amp_K0}
\end{equation}
where $p_{J/\psi}$ is the momentum of $J/\psi$ in the $\eta^\prime K^-$ rest frame, as follow,
\begin{equation}
	p_{J/\psi}=\dfrac{\lambda^{1/2}(M_{\eta^{\prime}K^-}^2,M_{B}^2,m^2_{J/\psi})}{2M_{\eta^{\prime}K^-}},
\end{equation}
and the $i=1,2,3$ correspond to the $\bar{K}^{*}\rho$, $\bar{K}^{*}\omega$, and $\bar{K}^{*}\phi$, respectively. According to Eq.~(\ref{eq:hadronize}), the weight coefficients are given as,
\begin{equation}
	h_{\bar{K}^{*}\rho}=\frac{3}{\sqrt{6}}, h_{\bar{K}^{*}\omega}=\frac{1}{\sqrt{2}}, h_{\bar{K}^{*}\phi}=1.
\end{equation}

We take the coupling constants $g_{K^{*}\rho}=(8102-i959)$~MeV, $g_{K^{*}\omega}=(1370-i146)$~MeV, $g_{K^{*}\phi}=-(1518+i209)$~MeV~\cite{Geng:2008gx}, and the two-meson loop function is given by 
\begin{equation}\label{G}
	G_i=i \int \frac{d^4 q}{(2 \pi)^4} \frac{1}{q^2-m_1^2+i \epsilon} \frac{1}{(P-q)^2-m_2^2+i \epsilon}.
\end{equation}
where $m_1$ and $m_2$ are the mesons masses of the $i$-th coupled channel. In the present work, we use the dimensional regularization method as indicated in Refs.~\cite{Duan:2020vye,Duan:2023qsg}, and in this scheme, the two-meson loop function $G_i$ can be expressed as,
\begin{equation}
	\begin{aligned}
		G_i= & \frac{1}{16 \pi^2}\left\{a(\mu)+\ln \frac{m_1^2}{\mu^2}+\frac{s+m_2^2-m_1^2}{2 s} \ln \frac{m_2^2}{m_1^2}\right. \\
		& +\frac{|\vec{q}\,|}{\sqrt{s}}\left[\ln \left(s-\left(m_2^2-m_1^2\right)+2 |\vec{q}\,| \sqrt{s}\right)\right. \\
		& +\ln \left(s+\left(m_2^2-m_1^2\right)+2 |\vec{q}\,| \sqrt{s}\right) \\
		& -\ln \left(-s+\left(m_2^2-m_1^2\right)+2 |\vec{q}\,| \sqrt{s}\right) \\
		& \left.\left.-\ln \left(-s-\left(m_2^2-m_1^2\right)+2 |\vec{q}\,| \sqrt{s}\right)\right]\right\},
	\end{aligned}
\end{equation}
here we take $\mu=1000$~MeV, and $a=-1.85$~\cite{Abreu:2023yvf,Geng:2008gx}. Since the vector mesons, particularly the $\rho$ and the $K^*$, are rather broad, one has to take into account their widths. We convolute the vector-vector $G$ function with the mass distributions of the two vector mesons, i.e., by replacing the $G$ function~\cite{Molina:2008jw,Ding:2023eps,Ding:2024lqk,Zhu:2022guw,Zhu:2022wzk,Lyu:2023aqn},
\begin{eqnarray}
	\tilde{G}(s)&=&\frac{1}{N^2}\int\limits^{(M_1+2\Gamma_1)^2}\limits_{(M_1-2\Gamma_1)^2}
	d\tilde{m}^2_1\left(-\frac{1}{\pi}\right)\mathrm{Im}\frac{1}{\tilde{m}^2_1-M_1^2+i\tilde{\Gamma}_1\tilde{m}_1}\nonumber\\
	&&\times\int\limits^{(M_2+2\Gamma_2)^2}\limits_{(M_2-2\Gamma_2)^2}
	d\tilde{m}^2_2\left(-\frac{1}{\pi}\right)\mathrm{Im}\frac{1}{\tilde{m}^2_2-M_2^2+i\tilde{\Gamma}_2\tilde{m}_2}
	\nonumber\\
	&&\times G(s,\tilde{m}_1^2,\tilde{m}_2^2)
\end{eqnarray}
with
\begin{eqnarray}
	N^2&=&\int\limits^{(M_1+2\Gamma_1)^2}\limits_{(M_1-2\Gamma_1)^2}
	d\tilde{m}^2_1\left(-\frac{1}{\pi}\right)\mathrm{Im}\frac{1}{\tilde{m}^2_1-M_1^2+i\tilde{\Gamma}_1\tilde{m}_1}\nonumber\\
	&&\times\int\limits^{(M_2+2\Gamma_2)^2}\limits_{(M_2-2\Gamma_2)^2}
	d\tilde{m}^2_2\left(-\frac{1}{\pi}\right)\mathrm{Im}\frac{1}{\tilde{m}^2_2-M_2^2+i\tilde{\Gamma}_2\tilde{m}_2},\nonumber
\end{eqnarray}
where $M_1$, $M_2$, $\Gamma_1$, and $\Gamma_2$ are the masses and
widths of the two vector mesons in the loop. We only take into
account the widths of the $\rho$ and the $K^*$. In the case of the $\omega$ or $\phi$ with narrow width, both the kernels of these integrals will reduce to a delta function $\delta(\tilde{m}^2-M^2)$. The $\tilde{\Gamma}_i$ function is energy dependent and has the form of
\begin{equation}
	\tilde{\Gamma}(\tilde{m})=\Gamma_0\frac{q^3_\mathrm{off}}{q^3_\mathrm{on}}\Theta(\tilde{m}-m_1-m_2)
\end{equation}
with
\begin{equation}
	q_\mathrm{off}=\frac{\lambda(\tilde{m}^2,m_\pi^2,m_\pi^2)}{2\tilde{m}},\quad
	q_\mathrm{on}=\frac{\lambda(M_\rho^2,m_\pi^2,m_\pi^2)}{2 M_\rho}
\end{equation}
and $m_1=m_2=m_\pi$ for the $\rho$ or
\begin{equation}
	q_\mathrm{off}=\frac{\lambda(\tilde{m}^2,m_K^2,m_\pi^2)}{2\tilde{m}},\quad
	q_\mathrm{on}=\frac{\lambda(M_{K^*}^2,m_K^2,m_\pi^2)}{2 M_{K^*}},
\end{equation}
$m_1=m_\pi$ and $m_2=m_K$ for the $K^*$, where $\lambda$ is the
K\"allen function, $\lambda(x,y,z)=(x-y-z)^2-4yz$, and $\Gamma_0$ is the nominal width of the $\rho$ or the $K^*$.

It is notable that the amplitude of Eq.~(\ref{eq:amp_K0}) does not satisfy the unitary requirements. Since we focus the invariant mass distribution and the intermediate resonances in this process, although the unitary condition of the amplitude is not ensured, one will see that our model can give a reasonable description of the experimental data in next section.

\subsection{Contribution of $Y(4710)$ in the $B^-\to J/\psi\eta^\prime K^-$}\label{sec2b}

\begin{figure}[htbp]
	\centering
	
	\includegraphics[scale=0.65]{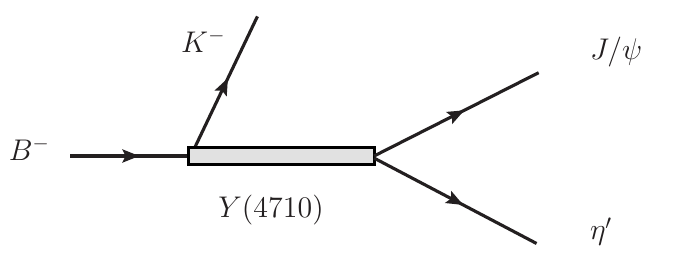}
	
	\caption{Mechanism for $Y(4710)$ state.}\label{fig:Y4710-hardon}
\end{figure}

The mechanism of $Y(4710)$ state is depicted in Fig.~\ref{fig:Y4710-hardon}, and the amplitude for $Y(4710)$ state is give by,
\begin{equation}
	\mathcal{T}^{Y(4710)}=\dfrac{V_{Y}\Tilde{k}_K\Tilde{k}_{\eta'}\text{cos}\Tilde{\theta}}{M_{J/\psi\eta^{\prime}}^2-M_{Y(4710)}^2+iM_{Y(4710)}\Gamma_{Y(4710)}},
\end{equation}
where the $V_{Y}$ is the relative weight with respect to the one of $K^*_0(1430)$, and the $\theta$ is angle between $J/\psi$ and $B^-$ in the rest frame of the $Y(4710)$~\cite{Wang:2015pcn,Wang:2022nac,Lyu:2024qgc,Zhang:2022xpf},
\begin{equation}
	\text{cos}\Tilde{\theta}=\dfrac{M_{ J/\psi K^-}^2-M_B^2-M_{\eta^{\prime}}^2+2P^0_BP_{\eta^{\prime}}^0}{2|\Tilde{k}_{\eta'}||\Tilde{k}_K|},
\end{equation}
\begin{equation}
	|\Tilde{k}_K|=\dfrac{\lambda^{1/2}(M_{J/\psi\eta^{\prime}}^2,M_{B}^2,m^2_{K^-})}{2M_{J/\psi\eta^{\prime}}},
\end{equation}
\begin{equation}
	|\Tilde{k}_{\eta '}|=\dfrac{\lambda^{1/2}(M_{J/\psi\eta^{\prime}}^2,m^2_{\eta^{\prime}},m^2_{J/\psi})}{2M_{J/\psi\eta^{\prime}}},
\end{equation}
and $P^0_B=\sqrt{M_B^2-(\tilde{k}_K)^2}$, $P^0_{\eta'}=\sqrt{m_{\eta'}^2-(\tilde{k}_{\eta'})^2}$. In this way we prepare the ground for expressing all the amplitudes in terms of the invariant masses, to finally calculate the mass distributions in terms of these invariant masses.

\subsection{Invariant Mass Distribution}\label{sec2c}

We can write the total amplitude of $B^-\to J/\psi\eta^\prime K^-$ reaction as,
\begin{equation}\label{Eq:t-total}
	\mathcal{T}=V_P\left\{\mathcal{T}^{K^*_0(1430)}+\mathcal{T}^{BG}e^{i\phi}+\mathcal{T}^{Y(4710)}e^{i\phi^\prime}\right\},
\end{equation}
where we introduce the background term from the non-resonant contribution,
\begin{equation}
	\mathcal{T}^{BG}=Cp_{J/\psi}, \label{eq:ampBG}
\end{equation}
and $V_P$ is a global normalization constant, and $C$, $\phi$ and $\phi^\prime$ are free parameters. Applying the standard formula of the RPP~\cite{ParticleDataGroup:2024cfk}, we have
\begin{equation}
	\frac{d^2\Gamma}{dM_{\eta^{\prime}K^-}dM_{J/\psi\eta^{\prime}}}=\frac{1}{(2\pi)^3}\dfrac{M_{\eta^{\prime}K^-}M_{J/\psi\eta^{\prime}}}{8M_{B}^3}|\mathcal{T}|^2.
\end{equation}

We can get $d\Gamma/dM_{12}$ by integrating $d^2\Gamma/(dM_{12}dM_{23})$ over $M_{23}$ with the limits in the RPP~\cite{ParticleDataGroup:2024cfk}. Permutation of the indices allows us to evaluate all three mass distributions, using $M_{12}$, $M_{23}$ as independent variables, and the property $M_{12}^2+M_{13}^2+M_{23}^2=M_{\Lambda_c^+}^2+m_{\pi^+}^2+m_\eta^2+M_\Lambda^2$ to get $M_{13}$ from them.

\section{Results and Discussions}\label{sec3}

\begin{figure}
	\subfigure[]{
		\includegraphics[scale=0.6]{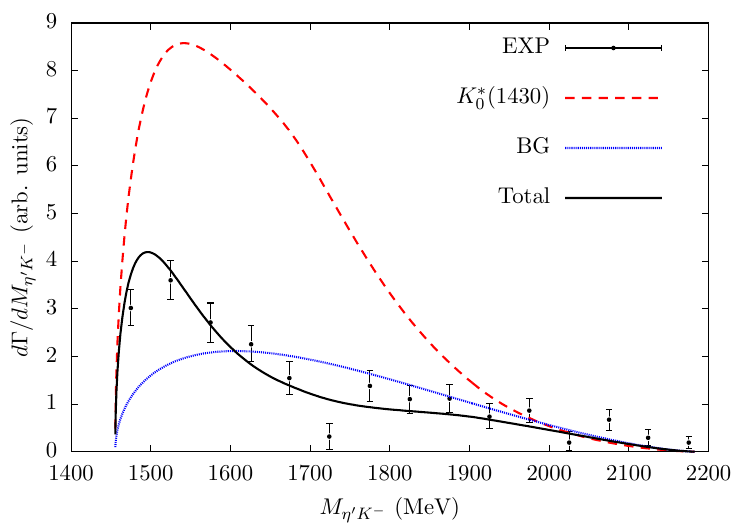}
	}
	\subfigure[]{
		\includegraphics[scale=0.6]{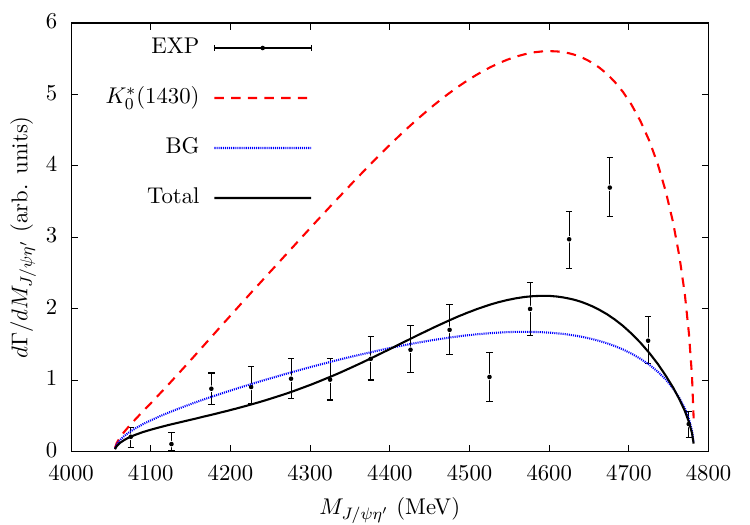}
	}
	\subfigure[]{
		\includegraphics[scale=0.6]{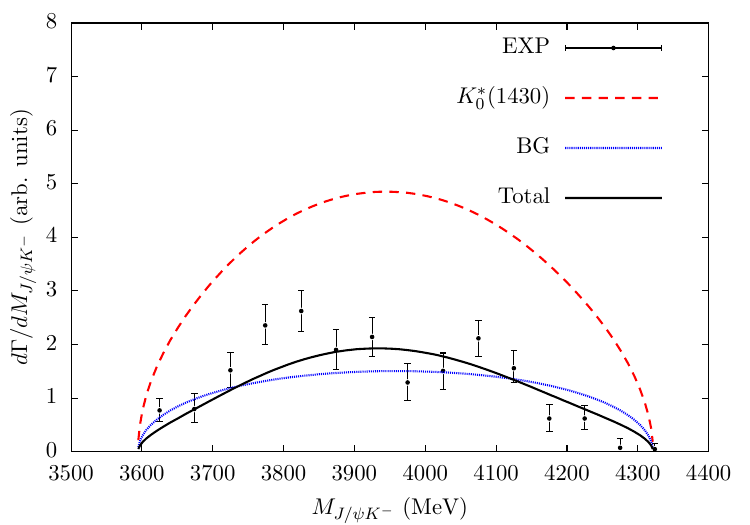}
	}
	\caption{Invariant mass distributions of $\eta^\prime K^-$~(a), $J/\psi\eta^\prime$~(b), and $J/\psi K^-$~(c) for the $B^-\to J/\psi\eta^\prime K^-$ reaction without the contribution of $Y(4710)$. The experimental data are taken from LHCb~\cite{LHCb:2023qca}.}\label{Fig:K+BG}
\end{figure}
\begin{figure}
	\subfigure[]{
		\includegraphics[scale=0.6]{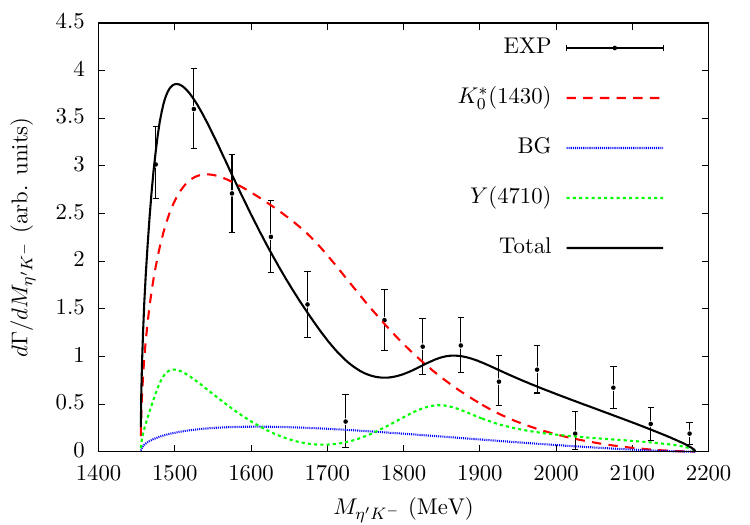}
	}
	\subfigure[]{
		\includegraphics[scale=0.6]{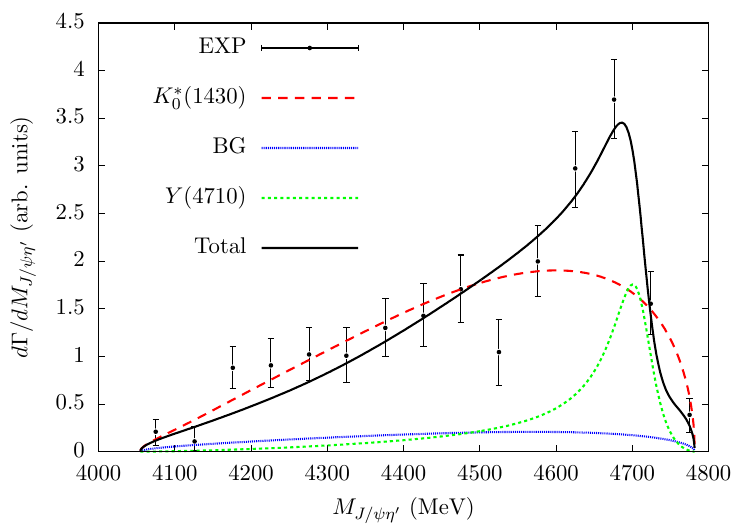}
	}
	\subfigure[]{
		\includegraphics[scale=0.6]{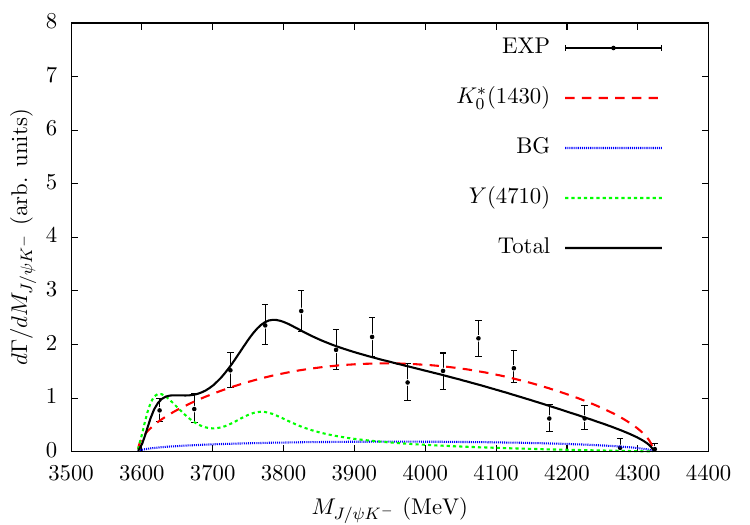}
	}
	\caption{Invariant mass distributions of $\eta^\prime K^-$~(a), $J/\psi\eta^\prime$~(b), and $J/\psi K^-$~(c) for the $B^-\to J/\psi\eta^\prime K^-$ reaction. The experimental data are taken from LHCb~\cite{LHCb:2023qca}.}\label{Fig:K+BG+Y}
\end{figure}

First, we have fitted to the LHCb invariant mass distributions of $\eta^\prime K^-$, $J/\psi\eta^\prime$, and $J/\psi K^-$ without the contribution of $Y(4710)$~\cite{LHCb:2023qca}. The fitted parameters are $V_P=1493.6$, $C=6.9932\times10^{-5}$, $\phi=-0.13118\pi$, and $\chi^2/d.o.f=108.6188/(44-3)=2.6492$. 
In Fig.~\ref{Fig:K+BG},  the red-dashed curves show the contribution of $\mathcal{T}^{K^*_0(1430)}$, the blue-dotted curves show the non-resonant contribution $\mathcal{T}^{BG}$ of Eq.~(\ref{eq:ampBG}), and the black solid curves show the contribution of total amplitude.
We observe at first glance that the agreement with the three experimental mass distributions is fair.
One can see that in the low energy region of the $\eta^\prime K^-$ mass distribution, this reaction is dominated by the $K_0^*(1430)$ state, and the interference between $K_0^*(1430)$ and background plays a key role. Although the basic features of the invariant mass distributions are reproduced, one can find that the peak structure in the high energy region of the $J/\psi\eta^\prime$ invariant mass distribution is not well reproduced.  Hence, we should consider the a charmonium-like state with mass around 4700~MeV and coupling to $J/\psi\eta^\prime$ channel.

Since the mass and width of the $Y(4710)$ have large experimental uncertainties, and we take them as free parameters obtained by fitting to the data of LHCb Collaboration~\cite{LHCb:2023qca}. Now, there are six free parameters and a global normalization constant. The fitted parameters are $V_p=870.12$, $C=4.2281\times10^{-5}$, $\phi=-0.25906\pi$, $V_{Y}=0.78186$, $\phi^\prime=1.5054\pi$, $M_{Y(4710)}=4710.7$~MeV, $\Gamma_{Y(4710)}=65.168$~MeV, and $\chi^2/d.o.f=58.64584/(44-7)=1.5850$, which is better than the previous one. The fitted mass and width of the $Y(4710)$ are in good agreement with the BESIII data within the uncertainties~\cite{BESIII:2022kcv,BESIII:2023wqy}. 

In Fig.~\ref{Fig:K+BG+Y}, we have  re-plotted the new results of the invariant mass distributions of $\eta^\prime K^-$, $J/\psi\eta^\prime$, and $J/\psi K^-$. The green-dotted curves show the contribution of $\mathcal{T}^{Y(4710)}$. In Fig.~\ref{Fig:K+BG+Y}(b), one can find that the peak structure in the high energy region of the $J/\psi\eta^\prime$ mass distribution can be well reproduced, which is mainly due to the $Y(4710)$ state. Furthermore, one can see that the  near-threshold structure in the $\eta^\prime K^-$ invariant mass distribution can be well described, and our prediction of the $J/\psi K^-$ invariant mass distribution is in good agreement with the LHCb data~\cite{LHCb:2023qca}, as shown in Figs.~\ref{Fig:K+BG+Y}(a) and \ref{Fig:K+BG+Y}(c), which supports the molecular explanation of the $K^*_0(1430)$,

\begin{figure}[htbp]
	\centering
	
	\includegraphics[scale=0.85]{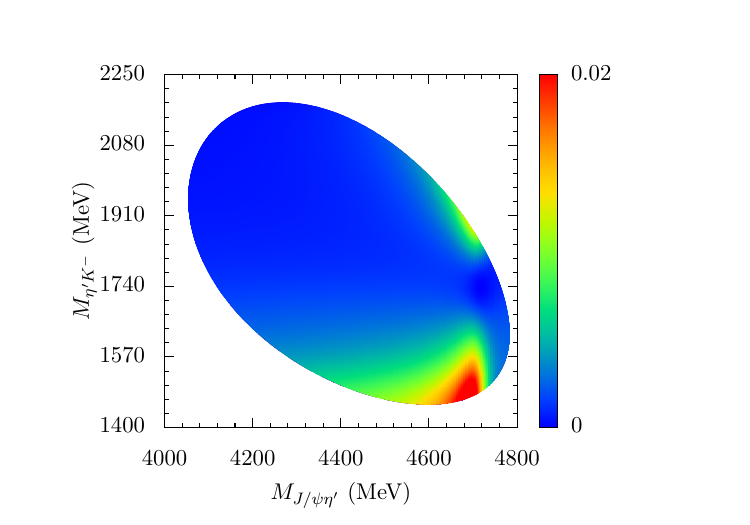}
	
	\caption{Dalitz plot of $J/\psi\eta^\prime$ vs. $\eta^\prime K^-$ for the $B^-\to J/\psi\eta^\prime K^-$ decay.}\label{fig:Dalitz}
\end{figure}

Then, we present the Dalitz plot for the $B^-\to J/\psi\eta^\prime K^-$ depending on $M_{J/\psi\eta^\prime}$ and $M_{\eta^\prime K^-}$ in Fig.~\ref{fig:Dalitz}. One can find that more events are expected to lie in the region of $M_{J/\psi\eta^\prime}>4600~\text{MeV}$ and $M_{\eta^\prime K^-}<1570~\text{MeV}$, in which a clear interference effect can be found, in agreement with the LHCb measurements~\cite{LHCb:2023qca}.


\section{ Conclusions }

Recently, the BESIII Collaboration has observed the charmonium-like state $Y(4710)$ with large uncertainties of the mass and width, and there are different  theoretical explanations for its structure. The LHCb Collaboration has analyzed the process $B^+\to J/\psi\eta^\prime K^+$ decay, and a interesting peak structure appears around 4700~Mev in the $J/\psi \eta^\prime$ invariant mass distribution, which could be associated with the charmonium-like state $Y(4710)$. 

In this work, we have carried out a detailed study of this process, where we consider the contributions from the $Y(4710)$ and the $K_0^*(1430)$. 
Firstly, we have calculated the invariant mass distributions without the $Y(4710)$, and found that our results cannot describe the peak structure around 4700~MeV in $J/\psi \eta^\prime$ invariant mass distribution. By taking account into the contribution of the $Y(4710)$ state, one can find that the peak structure around 4700~MeV in $J/\psi \eta^\prime$ mass distribution can be well reproduced. Furthermore, our results of the $\eta^\prime K^-$ and $J/\psi K^-$ invariant mass distributions are in good agreement with the LHCb measurements. Furthermore, we have plotted the Dalitz plot, which is also consistent with the LHCb results.

It it notable that the mass and width of the $Y(4710)$ state have large experimental uncertainties, and this state is not yet listed in RPP. Thus, it is very important to measure the properties of the $Y(4710)$ precisely in more processes. Meanwhile, if the $Y(4710)$ is a  charmonium state ($c\bar{c}$),
the decay mode of $Y(4710)\to J/\psi \eta'$ is OZI-suppressed, and the branching fraction of the mode $\mathcal{B}(Y(4710)\to J/\psi\eta')$ is expected to be very small.  If the measured branching fraction of this mode through the process $B^-\to J/\psi \eta' K^-$ or other processes is significantly larger than other decay modes, it could imply that the $Y(4710)$ could have the complex components, such as tetraquark component, in addition to the $c\bar{c}$ component. 
In all, we advocate that Belle~II and LHCb Collaborations could perform the more precise analysis to confirm the evidence of the $Y(4710)$ state~\cite{Jia:2023upb}.

\section*{Acknowledgments}

We acknowledge the support from the National Key R\&D Program of China (No. 2024YFE0105200 and No. 2024YFA1610503).
This work is supported by the Natural Science Foundation of Henan under Grant No. 232300421140, the National Natural Science Foundation of China under Grant No. 12475086, No. 12192263, No. 12175037, and No. 12335001. This work is supported by Zhengzhou University Young Student Basic Research Projects (PhD students) under Grant No. ZDBJ202522.

\end{document}